\documentstyle[eqsecnum,aps]{revtex}

\input epsf
\newcommand{\extraspace}{\addtolength{\abovedisplayskip}{2mm}
                        \addtolength{\belowdisplayskip}{2mm}
                        \addtolength{\abovedisplayshortskip}{2mm}
                        \addtolength{\belowdisplayshortskip}{2mm}}
\newcommand{\be}{\begin{equation}\extraspace}
\newcommand{\ee}{\end{equation}}
\newcommand{\bea}{\begin{eqnarray}\extraspace}
\newcommand{\eea}{\end{eqnarray}}
\newcommand{\nonu}{\nonumber \\[2mm]}

\newcommand{\th}{\theta}

\newcommand{\STR}{\rule[-5.5mm]{0mm}{13mm}}

\begin{document}

\title{\large Supersymmetric Reflection Matrices
  \footnote{Talk given at the International Seminar on
  {\em Supersymmetry and Quantum Field Theory}, dedicated
  to the memory of D.V.~Volkov, Kharkov (Ukraine), January
  5-7, 1997}
} 
\author{ M. Moriconi \footnote{{\small e.mail: moriconi@ictp.trieste.it}}} 
\address{ High Energy Section, ICTP\\
Strada Costiera 11, 34100 Trieste, Italy}

\author{K. Schoutens \footnote{{\small e.mail: kjs@phys.uva.nl}}}
\address{ Institute for Theoretical Physics, University of Amsterdam\\
Valckenierstraat 65, 1018 XE Amsterdam, The Netherlands}

\maketitle

\vskip 1cm

\begin{abstract}
We briefly review the general structure of integrable particle theories
in $1+1$ dimensions having $N=1$ supersymmetry. Examples are specific
perturbed superconformal field theories (of Yang-Lee type) and the
$N=1$ supersymmetric sine-Gordon theory. We comment on the modifications
that are required when the $N=1$ supersymmetry algebra contains non-trivial
topological charges.
\end{abstract}

\vfill

\noindent IC/97/34

\noindent ITFA-97-13

\newpage

\section{Introduction}

Quantum field theory (QFT) provides a powerful and unifying language to understand 
a variety of physical phenomena. In general we may define a QFT by choosing a set 
of fields that transform according to some irreducible representation of the 
Poincar\'e group, together with a prescription (for example, coming from
a lagrangian) that gives us the dynamics of these fields. In the context of 
Particle Physics, it is strongly believed that the Poincar\'e group is a true 
symmetry of the world and therefore we always take it for granted. 

In general, a QFT becomes more tractable if it possesses additional symmetries
beyond Poincar\'e invariance. Via Ward identities, extra symmetries provide strong
constraints on correlation functions and so make possible a more thorough analytical
treatment. To introduce new symmetries in a QFT is rather delicate: we do not
want to oversimplify the specific models we are looking at but we would like to 
have enough symmetry to improve the physical properties and to gain the upper hand in 
controlling the theory. Among the possible symmetries we may consider, 
{\em supersymmetry}\ stands out as a very special one. Supersymmetry unifies the 
apparently incompatible concepts of bosons and fermions, and often improves the 
physical properties of specific field theories. At the same time, supersymmetric 
theories are usually easier to analyze. This last remark applies to supersymmetric 
Yang-Mills theory (in 4 dimensions), to string models (in 10 dimensions) and to 
models of QFT in 2 dimensions.

Focussing on QFT's in $1+1$ dimensions, we may further specialize to models
that are {\em integrable}. By this we mean that we assume the existence of 
`enough' (meaning an infinite number of) charges in involution.%
\footnote{To say that charges are in involution means (in classical mechanics) that
  the Poisson brackets of any two of them vanish or (in QFT) that they all mutually
  commute.}
For a QFT describing (massive) particles, integrability implies the factorizability 
of the scattering matrix: the S-matrix for $n$-particle scattering factorizes 
into two-body S-matrices, there is no particle production and the individual momenta 
of the particles are conserved. By assuming both supersymmetry and integrability,
we thus arrive at very simple supersymmetric particle theories which can be
analyzed in closed form, and which may serve as prototypes for supersymmetric
particle theories in higher dimensions.

In addition to the motivation we have given so far, there are more direct
reasons for considering QFT's in two dimensions. For one thing, such theories
directly apply to the analysis of either classical systems of statistical mechanics 
in 2 dimensions, or quantum mechanical systems in 1+1 dimensions (that is, on a 
line). In addition, there are applications to problems in $3+1$ dimensions, where 
the essential physics takes place in the radial direction. In the latter type of 
applications, the models live on a half-line and the behavior at the
boundary%
  \footnote{the origin of space in the original three-dimensional formulation}
is important. Examples are 
the Callan-Rubakov effect (the catalysis of baryon decay in the field of a 
magnetic monopole),
the Hawking effect (quantum black-hole evaporation), 
the Kondo problem (magnetic impurities coupling to conduction electrons) 
or edge current tunneling in the quantum Hall effect. 

In view of these applications, it is an interesting problem to consider
integrable supersymmetric particle theories in $1+1$ dimensions in the presence
of a boundary. In a recent paper \cite{MS2} we obtained a general form for boundary
reflection matrices in $N=1$ supersymmetric theories, and we worked out a
number of examples.

In this note we outline the general structure of integrable supersymmetric
QFT's in $1+1$ dimensions, paying particular attention to their boundary scattering. 
We shall briefly introduce specific examples (which are perturbations of 
supersymmetric Yang-Lee-type conformal field theories and the breathers in the 
supersymmetric sine-Gordon theory). We shall also 
comment on the extension of these results to the case of $N=1$ supersymmetry 
with non-zero topological charge.

\section{S-matrices and Reflection Matrices: General}

Given a bulk integrable field theory, one may start the analysis by determining
the two-body scattering matrix. This is a key ingredient for the understanding
of the physics of the model and the first step towards computing
correlation functions using the form-factor approach. From now on the term 
``S-matrix" will be used for the two-body S-matrix unless stated otherwise 
explicitly. We will use the rapidity variable $\theta$, which parametrizes 
the on-shell momenta of the particles by $p_0=m\cosh(\theta)$ and 
$p_1=m\sinh(\theta)$. The S-matrix between particles 1 and 2 can be written 
as $S_{a_1a_2}^{b_1b_2}(\theta_{12}))$, where $\theta_{12}=\theta_1-\theta_2$ 
is the difference of the rapidities of the incoming particles. 

Let us now briefly outline the general strategy for obtaining the S-matrix 
for an integrable model with some non-trivial symmetries. One starts by 
writing down the most general S-matrix compatible with the unbroken symmetries 
and then requires integrability. This is done by imposing the famous
Yang-Baxter equation (YBE). The YBE is shown in figure 1.

\vskip 0.5cm
\centerline{\epsffile{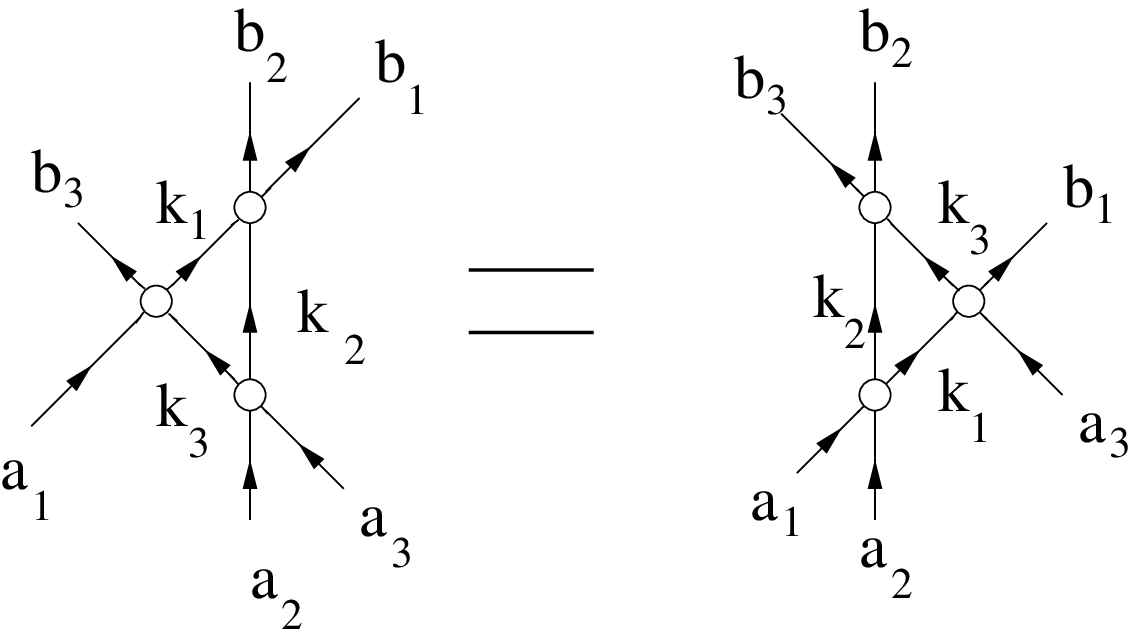}}
\vskip 0.5cm
\centerline{Fig. 1 Pictorial representation of the YBE}
\vskip .3cm
In a formula the YBE reads
\bea
\sum_{k} S_{a_1a_2}^{k_1k_2}(\theta_{12}) S_{k_1a_3}^{b_1k_3}(\theta_{13})
S_{k_2k_3}^{b_2b_3}(\theta_{23})= \nonu
=\sum_{k} S_{a_2a_3}^{k_2k_3}(\theta_{23})
S_{a_1k_3}^{k_1b_3}(\theta_{13})
S_{k_1k_2}^{b_1b_2}(\theta_{12}) \ . \
\eea

Once the YBE is solved we impose the usual constraints from general S-matrix theory,
that is, analyticity, crossing-symmetry and unitarity. After we managed to do all
that (it can be done in many cases!) we have to impose the so-called {\em bootstrap
principle}: bound states are to be treated on the same footing as asymptotic states.
This, together with integrability as encoded in the YBE, provides a very restrictive 
set of equations that greatly constrain the initial S-matrix. Once we reach a 
self-consistent spectrum we will have found the minimal S-matrix for our model. Of 
course this can not be the whole story, since different models with the same symmetry 
and same spectrum may correspond to quite different lagrangians, say. This ambiguity 
is indeed present and it is called CDD ambiguity, after the work of Castillejo, Dalitz
and Dyson \cite{CDD}. 

One of the ways to test a conjectured S-matrix is through the
thermodynamic Bethe Ansatz (TBA). This is a general procedure where we start with
the S-matrix as input and compute some ultraviolet physical properties such as the 
ground state energy (central charge) and scaling dimensions of the underlying QFT. 
Comparing these with ultraviolet data obtained from a lagrangian or from conformal 
field theory, we have a non-trivial check on the conjectured S-matrix.

Next we go from the bulk theory to a theory defined on half-line. We will have to
specify the boundary action or simply assume that the boundary action is such as to
preserve integrability and the extra symmetries of the theory. The theory is then
described in the bulk by the same S-matrix as before but now we have to find 
reflection matrices $R_a^b(\theta)$, which tell us how particles scatter off the 
boundary. We will assume that the boundary does not change the particle species,
so that $R_a^b(\theta)=\delta_a^b R_a(\theta)$. Integrability is imposed now via 
the boundary Yang-Baxter equation (BYBE), which can be represented in this case as

\vskip 0.5cm
\centerline{\epsffile{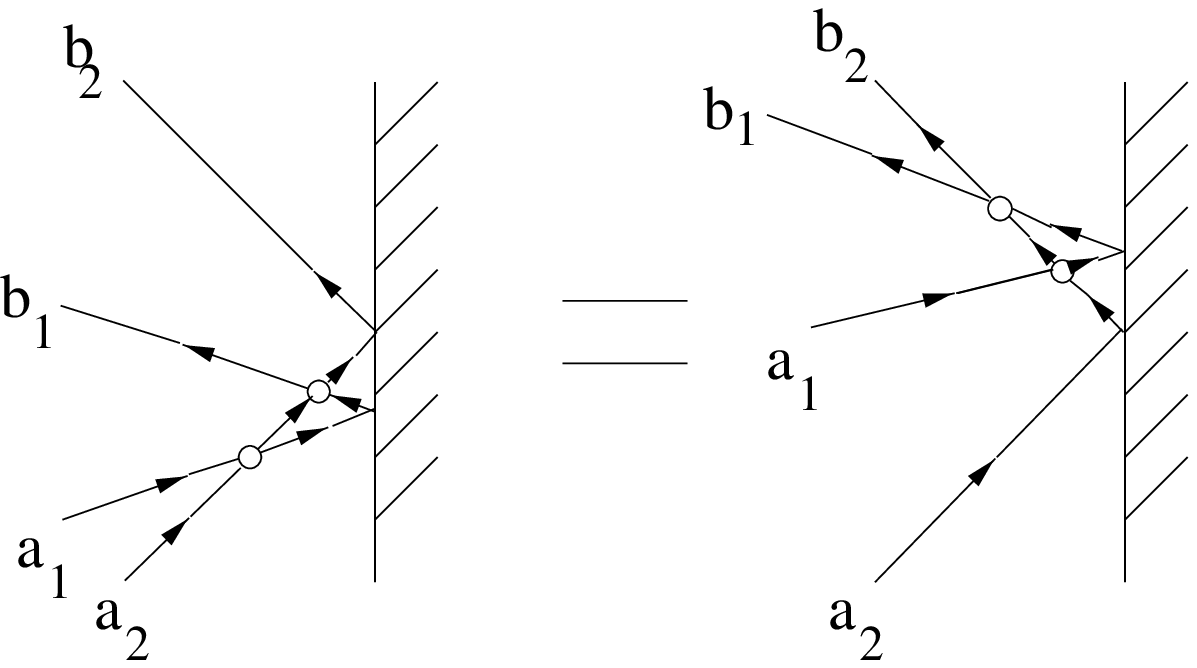}}
\vskip 0.5cm
\centerline{Fig. 2 Pictorial representation of the BYBE}
\vskip .3cm
In a formula%
   \footnote{no sum over $a_1,a_2,b_1,b_2$},
\bea
\lefteqn{
R_{a_2}(\th_2)
S_{a_1a_2}^{c_1d_2}(\th_1+\th_2)
R_{c_1}(\th_1)
S_{d_2c_1}^{b_2b_1}(\th_1-\th_2)=}
\nonu
&& \qquad
S_{a_1a_2}^{c_1c_2}(\th_1-\th_2)
R_{c_1}(\th_1)
S_{c_2c_1}^{b_2b_1}(\th_1+\th_2)
R_{b_2}(\th_2) \ .
\label{bdr}   
\eea

The general idea is clear now. Once we found an exact (bulk) S-matrix we can study
boundary versions of the same theory by solving the BYBE and the analogous
requirements on the reflection matrix (analyticity, unitarity, boundary
crossing-unitarity). Note that the introduction of a boundary necessarily changes
the structure of conserved charges. Some of the charges that were conserved in the 
bulk are not conserved anymore, e.g., linear momentum.

\section{$N=1$ Supersymmetry Without Topological Charges}

Supersymmetry may sound as an odd concept in $1+1$ dimensions, since the very 
definition
of bosons and fermions is based on the behavior of the matter field under the
rotation group. Since we do not have a rotation group in $1+1$ dimensions one 
may feel
uneasy with such concepts. One may realize, however, that the Lorentz group
(which is non-trivial in 1+1 dimension) suffices to define the notions
of bosonic fields (of integer Lorentz spin) and fermionic fields
(of half-integer Lorentz spin). Supersymmetry comes with a parity operator
$Q_L$, which has eigenvalue $+1$ on bosonic states and $-1$ on
fermionic states. The $N=1$ supersymmetry algebra takes the form 
\bea 
&& \qquad \qquad \{Q_L,Q_{\pm}\}=0 \nonu
&&Q_+{}^2=p_0+p_1, \qquad Q_-{}^2=p_0-p_1 \\[2mm] 
&& \qquad \qquad \{Q_+,Q_-\}=0\ . 
\nonumber 
\eea 
The anticommutator $\{Q_+,Q_-\}$ could have been non-zero, and equal to a real
number $Z \leq 2$, which would correspond to a topological charge $Z$. We will
initially consider the situation $Z=0$ and later show how it can be generalized to
$Z \neq 0$. We will use the following realization
\be
Q_+(\theta)= \sqrt{m} \, e^{\theta \over 2}
\left(\begin{array}{cc}
      0 & 1\\
      1 & 0
      \end{array}\right), \quad
Q_-(\theta)= \sqrt{m} \, e^{-{\theta \over 2}}
\left(\begin{array}{cc}
      0 & -i\\
      i & 0
      \end{array}\right), \quad 
Q_L(\theta)=
\left(\begin{array}{cc}
      1 & 0\\
      0 & -1
      \end{array}\right)\ \ . \ 
\ee
We can then define how these operators act on multi-particle states
(all we need are two-particle states) and impose that the
S-matrix commutes with the supersymmetry generators \cite{Sch}. There is
one point here that we should stress: to date all exact $N=1$ S-matrices
are of a special form, given by
\be
S^{[ij]}(\theta)=S^{[ij]}_{B}(\theta)S^{[ij]}_{BF}(\theta) \ , \ \label{smatrix}
\ee
where $S^{[ij]}_{B}(\theta)$ is the S-matrix of the bosonic projection of the
theory and $S^{[ij]}_{BF}(\theta)$ is a universal S-matrix that mixes bosons and 
fermions in such a way that the final non-diagonal S-matrix commutes with the
supersymmetry charges. The bosonic factor $S^{[ij]}_{B}(\theta)$,
describing the scattering of bosons $b_i$ and $b_j$, is assumed
to be diagonal. All the physical (bound-state) poles of the total S-matrix
are contained in this factor. The fermions are labeled by $f_i$ with $i=1,2,\ldots, n$.
The particles $b_i$ and $f_i$ have the same mass and form a supermultiplet
under the $N=1$ supersymmetry. The general $S^{[ij]}_{BF}$ matrix was proposed by 
one of us in \cite{Sch} and we refer to that paper for a more complete discussion. 

In \cite{Sch} it was found that integrability and supersymmetry alone fix the form
of $S^{[ij]}_{BF}(\theta)$ up to one constant $\alpha$. To fix this constant we
have to look at the bootstrap relations. It can be shown that if the particles 
$b_i$ and $b_j$ for a bound state $b_k$ then we have the following relation
\be
\alpha=-{{(2 m_i^2 m_j^2 + 2 m_i^2 m_k^2 + 2 m_j^2 m_k^2-
m_i^4-m_j^4-m_k^4)^{1 \over 2}} \over {2 m_i m_j m_k}} \ . \ \label{alpha}
\ee
Another consequence of supersymmetry is that once we have a three-point coupling
between particles $b_i$, $b_j$ and $b_k$ we will also have three-point couplings for
($f_i$, $f_j$, $b_k$), ($f_i$,$b_j$,$f_k$) and ($b_i$,$f_j$,$f_k$). The following
ratio is then obtained 
\be
{{f_{f_i f_j b_k}} \over {f_{b_i b_j b_k}}}=\left( {m_i+m_j-m_k} \over {m_i+m_j+m_k}
\right)^{1 \over 2} \ . \
\ee

{}From conditions such as (\ref{alpha}) it is clear that bosonic theories that
can be supersymmetrized in this simple manner have to be rather special.
In all known examples, the masses $m_i$ of the particles $b_i$ come out
as 
\be
m_i={{\sin(i \beta \pi)} \over {\sin(\beta \pi)}}, 
\qquad \beta={1 \over {2n+1}} \ , \ \label{masses}
\ee
for $i=1,2, \ldots, n$, $\alpha=-\sin(\beta \pi)$, and there is a specific bound
state structure, related to $A_{2n}^{(2)}$ group theory. In section V we shall
present some explicit examples.

\section{Supersymmetric Reflection Matrices}

In this section we will explain how to obtain boundary reflection matrices for $N=1$
supersymmetric theories by circumventing some of the typical difficulties of boundary
integrable models. The basic assumption is that the boundary action is such that
{\em integrability and supersymmetry are both preserved}. On top of that we assume
that the reflection matrix can be factorized in a similar fashion to the bulk
S-matrix,
\be
R(\theta)=R_{B}(\theta) \, R_{BF}(\theta) \ . \ \label{rmatrix}
\ee
The $R_{B}(\theta)$ factor is the reflection matrix for the bosonic projection of
the theory, and the $R_{BF}(\theta)$ is the ``supersymmetric" piece.
This factor has the following representation in a $|b\rangle$, $|f\rangle$ basis
\be
R_{BF}(\theta)=
\left(\begin{array}{cc}
      R_{bb}(\theta) & R_{bf}(\theta)\\
      R_{fb}(\theta) & R_{ff}(\theta)
      \end{array}\right) . \quad
\ee
As we will see now, this
will be enough to fix almost completely the reflection matrices, in a similar way to
what happens in the bulk case.

If we assume that supersymmetry is preserved by the boundary action we have to
impose the ``commutation" relation between the reflection matrix and some linear
combination of the two bulk supercharges
\be
{\cal Q}(\theta)R(\theta)=R(\theta){\cal Q}(-\theta) \ , \ \label{commutator}
\ee
where ${\cal Q}(\theta)=aQ_+(\theta)+bQ_-(\theta)$, $a$ and $b$ some arbitrary
real numbers. It is easy to see that the only solutions for (\ref{commutator})
are $a=\pm b$ and
\be
R_{BF}^{\pm}(\theta)=Z^{\pm}(\theta)
\left(\begin{array}{cc}
      \cosh({\theta \over 2} \pm i{\pi \over 4}) & e^{i{\pi \over 4}}Y(\theta)\\
      e^{-i{\pi \over 4}}Y(\theta) & \cosh({\theta \over 2} \mp i{\pi \over 4})
      \end{array}\right). \quad
\ee
By imposing BYBE we find that $Y(\theta)=0$. This means that the boundary can not
change the fermion number of the incoming particle.
   
The mass (supermultiplet) dependence of $R_{BF}^{(\pm)}$ is encoded in the prefactor
$Z^{(\pm)}$. So in order to have a complete description of boundary scattering we
have to fix these functions, by imposing unitarity and boundary crossing-symmetry. 
This was done in \cite{MS2} and we refer to that paper for details. On the other
hand, without any further work we can see immediately that the 
ratio $R_b/R_f$ is universal
\be {{R_b^{\pm}(\theta)} \over {R_f^{\pm}(\theta)}}={{\cosh({\theta \over 2}
\pm i{\pi \over
4})} \over {\cosh({\theta \over 2} \mp i{\pi \over 4})}} \ . \ \label{rbf}
\ee
These results follows directly from supersymmetry, the factorization Ansatz
(\ref{rmatrix}) and the specific realization of the superalgebra that we are using.

\section{Examples: Supersymmetric Yang-Lee Models and Breathers
in the Supersymmetric sine-Gordon Model}

In this section we give some examples of theories that realize the general
structure presented in sections III and IV.

The first series of examples are the so-called supersymmetric generalized
Yang-Lee models. They are obtained as integrable deformations of specific
$N=1$ superconformal field theories of central charges $c=-3n(4n+3)/(2n+2)$, 
$n=1,2, \ldots$, where the perturbing field is the bottom component of 
the Neveu-Schwarz field labeled as $\phi_{(1,3)}$. The spectrum of the 
massive deformation is as in (\ref{masses}). The first model of this series 
corresponds to the supersymmetrization of the Yang-Lee model. 

The supersymmetric sine-Gordon theory is defined by the following action
in euclidean space-time
\bea
S_{ssG}=\int_{-\infty}^{\infty}dy\int_{-\infty}^{\infty}dx  
&&\left\{ \STR {1\over2}(\partial_x\phi)^2+{1\over2}(\partial_y\phi)^2-
{\bar{\psi}}(\partial_x-i\partial_y){\bar{\psi}}+
\psi(\partial_x+i\partial_y)\psi-\right.\nonu
&&\left.-{{m^2}\over{\beta_{ssG}^2}}\cos(\beta_{ssG}\phi)-
2m{\bar{\psi}}\psi\cos({{\beta_{ssG}\phi}\over2}) \STR \right\} , 
\label{ssG1}
\eea
where $\phi$ is the bosonic field and $\psi$ and $\bar{\psi}$ are the
components of a Majorana fermion. The spectrum of the full quantum theory 
contains (anti-)soliton multiplets and bound state multiplets
$(b_j,f_j)$, $j=1,2,\ldots < \lambda$, $\lambda = 
2\pi \left( 1 - (\beta_{ssG}^2 / 4\pi) \right)/\beta_{ssG}^2 $,
of masses (\ref{masses}) with $\beta= 1/(2\lambda)$.

In all these examples, the bulk S-matrices and boundary R-matrices are
of the general form discussed in sections III and IV \cite{Sch,Ahn1,MS2}. 
The detailed form of the reflection matrices was worked out in our recent 
paper \cite{MS2}.

We already mentioned that the thermodynamic Bethe Ansatz (TBA) 
is a very effective way to test the validity of conjectured S-matrices.
While this analysis is more or less routine for diagonal S-matrices, 
the analysis for non-diagonal S-matrices is non-trivial and has
to be studied on a case by case basis. Fortunately, $N=1$ supersymmetric 
integrable models can be mapped into the eight-vertex model at a special 
point, where they satisfy the so-called ``free-fermion" condition, which 
allows to complete the TBA program.  This was done in \cite{Ahn} for the 
super Yang-Lee case and in \cite{MS1} for the more general perturbed 
superconformal field theories discussed in this section.

In the case of the supersymmetric sine-Gordon theory, we have been able
to propose exact reflection matrices without knowing the boundary
action. An interesting problem is then to find the boundary actions that 
correspond to these matrices. Inami, Odake and Zhang \cite{IOZ} have proposed 
two possible boundary actions that preserve integrability and supersymmetry.
Their proposal is based on the study of the conserved charges, at the
classical level, in the presence of a boundary. In \cite{MS2} we 
established a connection between this proposal and our reflection matrices 
by looking at the weak coupling limit of the supersymmetric sine-Gordon model 
\cite{MS2}.

\section{Topological Charges}

In the presence of topological charges, the anticommutator $\{Q_+,Q_-\}$
changes to
\be
\{Q_+,Q_-\}= Z \ , \
\ee
with $Z \leq 2$.%
\footnote{This $Z$ should not be confused with the
   prefactors $Z^{(\pm)}$.} 
We will adopt the following realization
\be
Q_+(\theta)= \sqrt{m} \, e^{\th \over 2}
\left(\begin{array}{cc}
      0 & 1\\
      1 & 0
      \end{array}\right), \quad
Q_-(\theta)= \sqrt{m} \, e^{-{\th \over 2}}
\left(\begin{array}{cc}
      0 & {\rm e}^{i \alpha}\\
      {\rm e}^{-i \alpha} & 0
      \end{array}\right), \quad
Q_L(\theta)= 
\left(\begin{array}{cc}
      1 & 0\\
      0 & -1
      \end{array}\right)\ \ , \
\ee
where $\cos(\alpha)=Z/2$. Note that the case $Z=0$ is obtained when
$\alpha=-\pi/2$.  Similarly to the case without topological charges we
assume that the reflection matrices will be of the same factorized form as
in (\ref{rmatrix}). 

Following the same approach as in the case without topological charges it is easy
to see that the ``supersymmetric'' part of the reflection matrix will have the
following form
\be 
R^{(\pm)}_{BF}(\theta)=Z^{(\pm)}(\theta)
\left(\begin{array}{cc}
     \cosh({\theta \over 2}+i({\pi \over 4}+{\alpha \over 2})
     \pm i{\pi \over 4}) & 0\\
       0 &  \cosh({\theta \over 2}+i({\pi \over 4}+{\alpha \over 2})
     \mp i{\pi \over 4})
      \end{array} \right)\ \ . \label{Rtopo} \
\ee
At $Z=0$ this reduces to the reflection matrix in
(\ref{rbf}). Again we recall that this is the reflection matrix obtained
by imposing that the boundary action preserves supersymmetry and
integrability. Notice that we again have a universal ratio
\be
{{R_b^{\pm}} \over {R_f^{\pm}}}={{\cosh({\theta \over 2}+
        i({\pi \over 4}+{\alpha \over 2})  
     \pm i{\pi \over 4})} \over {\cosh({\theta \over 2}+
        i({\pi \over 4}+{\alpha \over 2})
     \mp i{\pi \over 4})}} \ . \
\ee

Recently, Hollowood and Mavrikis \cite{HM} have proposed exact $N=1$
supersymmetric S-matrices for theories with non-zero topological 
charges. We expect that the reflection matrices (\ref{Rtopo}) can 
consistently be combined with these new $S$-matrices, in the sense that 
together they form a solution of the BYBE.

\section{Acknowledgements}

We would like to thank Roland K\"oberle and Andreas Fring for useful 
discussions. One of us (MM) would like to thank the University of Amsterdam, 
where part of this work was done, for the warm hospitality. The research
of KS was supported in part by the foundation FOM of the Netherlands.

\end{document}